\documentclass[a4paper,fleqn,usenatbib,useAMS]{mnras}

\usepackage[T1]{fontenc}

\usepackage[english]{babel}
\usepackage{graphicx}
\usepackage{fleqn}
\usepackage{amssymb}
\usepackage{amsmath}
\usepackage{booktabs}
\usepackage{hyperref}
\usepackage{enumitem}
\renewcommand{\S}{Section}
\newcommand{\F}{Fig.}

\title[White dwarf pollution in binaries]{White dwarf pollution by planets in stellar binaries}
\author[Adrian S. Hamers and Simon F. Portegies Zwart]{Adrian S. Hamers$^{1}$ and Simon F. Portegies Zwart$^{1}$ \\
$^{1}$Leiden Observatory, Leiden University, PO Box 9513, NL-2300 RA Leiden, The Netherlands}

\pubyear{2016}
\date{Accepted 2016 July 05. Received 2016 June 30; in original form 2016 June 17}

\begin{document}
\label{firstpage}
\pagerange{\pageref{firstpage}--\pageref{lastpage}}
\maketitle

\begin{abstract} 
Approximately $0.2 \pm 0.2$ of white dwarfs (WDs) show signs of pollution by metals, which is likely due to the accretion of tidally disrupted planetary material. Models invoking planet-planet interactions after WD formation generally cannot explain pollution at cooling times of several Gyr. We consider a scenario in which a planet is perturbed by Lidov-Kozai oscillations induced by a binary companion and exacerbated by stellar mass loss, explaining pollution at long cooling times. Our computed accretion rates are consistent with observations assuming planetary masses between $\sim 0.01$ and $1\,M_\mathrm{Mars}$, although nongravitational effects may already be important for masses $\lesssim 0.3 \, M_\mathrm{Mars}$. The fraction of polluted WDs in our simulations, $\sim 0.05$, is consistent with observations of WDs with intermediate cooling times between $\sim 0.1$ and 1 Gyr. For cooling times $\lesssim 0.1$ Gyr and $\gtrsim 1$ Gyr, our scenario cannot explain the high observed pollution fractions of up to 0.7. Nevertheless, our results motivate searches for companions around polluted WDs. 
\end{abstract}


\section{Introduction}
\label{sect:introduction}

The atmospheres of cool white dwarfs (WDs) are expected to consist entirely of hydrogen or helium due to efficient gravitational settling of metals \citep{1945AnAp....8..143S}. However, in $0.2 \pm 0.2$ of white dwarfs \citep{2006A&A...453.1051K,2014A&A...566A..34K}, spectra have revealed emission lines from a large range of metals, suggesting that these `polluted' WDs have recently accreted metal-rich material (see \citealt{2014AREPS..42...45J,2016RSOS....3.0571V,2016NewAR..71....9F} for reviews). Observations indicate that the pollution rate is approximately independent of cooling time \citep{2014A&A...566A..34K}, requiring a continuous pollution process.

Accretion from the interstellar medium \citep{1993ApJS...87..345D} has been ruled out \citep{2003ApJ...596..477Z,2006A&A...453.1051K,2007ApJ...663.1291D,2008AJ....135.1785J}. WD pollution could instead originate from accreting tidally disrupted rocky planetary material (e.g. \citealt{1986ApJ...302..462A,1993AJ....105.1033A,2002ApJ...572..556D,2003ApJ...584L..91J}) with a composition similar to Earth's (see e.g. \citealt{2014AREPS..42...45J}, and references therein), originating from planetesimals of mass $\sim 10^{20}\,\mathrm{kg}$ to planets as massive as Mars \citep{2009ApJ...699.1473J}. This is supported by the observation that all WDs with discs are polluted, and by the observed transiting planetesimals in tight orbits around WD 1145+017 \citep{2015Natur.526..546V}.

Polluted WDs are therefore a probe for planetary systems around WDs (see \citealt{2016RSOS....3.0571V} for a review). Bodies in tight orbits are engulfed by the star as it expands along the red giant branch (RGB; \citealt{2009ApJ...705L..81V,2011ApJ...737...66K,2014ApJ...794....3V}) and asymptotic giant branch (AGB; \citealt{2012ApJ...761..121M}) phases. At larger distances, stellar mass loss, tides, interactions with stellar ejecta and nongravitational effects are important. Early after WD formation, dynamical instabilities arising from planet-planet interactions and mass loss could lead to the disruption of planetary material and WD pollution \citep{2002ApJ...572..556D,2011MNRAS.414..930B,2012ApJ...747..148D,2016MNRAS.458.3942V}. These instabilities typically occur on short time-scales, and cannot explain continued pollution of WDs with cooling times of several Gyr. 

\citet{2015MNRAS.454...53B} proposed a scenario independent of the WD cooling time, in which the WD planetary system is perturbed by a wide binary companion whose orbit is driven to high eccentricity due to Galactic tides.

We investigate a related scenario in which the WD and planet are orbited by a secondary star. We focus on planets with radii $\gtrsim 1000 \, \mathrm{km}$, for which nongravitational effects are not important (e.g. \citealt{2016RSOS....3.0571V}). Mass loss of the primary star triggers adiabatic expansion of both the inner (planet's) and outer (secondary's) orbits. The importance of Lidov-Kozai (LK) oscillations \citep{1962P&SS....9..719L,1962AJ.....67..591K} in the inner orbit then typically increases \citep{2012ApJ...760...99P,2013ApJ...766...64S,2013MNRAS.430.2262H,2014ApJ...794..122M}. Consequently, the inner orbit can be driven to high eccentricity for the planet to be tidally disrupted by the WD, polluting the latter. Pollution can be prolonged to several Gyr after the WD formed.

\section{Methodology}
\label{sect:methodology}
\subsection{Algorithm}
\label{sect:methodology:alg}
We used the secular dynamics code of \citet{2016MNRAS.459.2827H} coupled with the stellar evolution code \textsc{SeBa} \citep{1996A&A...309..179P,2012A&A...546A..70T}. In \textsc{SeBa}, we assumed a metallicity of 0.02. Adiabatic mass loss was assumed to compute the dynamical response of the orbits on mass loss. Tidal evolution was modelled with the equilibrium tide model \citep{1998ApJ...499..853E}. For the primary star, the tidal dissipation strength was computed using the prescription of \citep{2002MNRAS.329..897H} with an apsidal motion constant of 0.014, a gyration radius of 0.08, an initial spin period of 10 d and zero obliquity (similar to \citealt{2007ApJ...669.1298F}). The stellar spin period was computed assuming conservation of spin angular momentum. For the planet, we assumed a viscous time-scale of $\approx 1.4 \, \mathrm{yr}$ \citep{2012arXiv1209.5724S}, an apsidal motion constant of 0.25, a gyration radius of 0.25, an initial spin period of $10 \, \mathrm{hr}$ and zero obliquity.

\subsection{Initial conditions}
\label{sect:methodology:IC}
$N_\mathrm{MC}=10^5$ systems were generated as follows. The primary mass $M_\star$ was sampled from a Salpeter distribution \citep{1955ApJ...121..161S} between 1.2 and 6 $M_\odot$. The secondary mass $M_\mathrm{c}$ was sampled assuming a linear distribution of $q = M_\mathrm{c}/M_\star$ with $0.1<q<1$. The mass of the planet, $m_\mathrm{p}$, was sampled logarithmically between $0.3\, M_\mathrm{Mars}$ and $1 \, M_\mathrm{J}$. The planetary radius was computed using the mass-radius relation of \citet{2013ApJ...768...14W}. According to the latter relation, $0.3\, M_\mathrm{Mars}$ corresponds to $\approx 1000 \, \mathrm{km}$.

We focused on planets with initial semimajor axes $a_1>1\,\mathrm{AU}$, for which interactions with stellar ejecta can be neglected. A linear distribution of $a_1$ was assumed between 1 and 100 AU. The outer orbit semimajor axis $a_2$ was sampled assuming a lognormal distribution of the outer orbital period between 10 and $10^{10}$ d \citep{1991A&A...248..485D,2010ApJS..190....1R,2014AJ....147...86T}. The eccentricities $e_i$ were sampled from a Rayleigh distribution with an rms of 0.33 \citep{2010ApJS..190....1R}. The orbits were assumed to be randomly orientated. A sampled configuration was rejected if the stability criterion of \citet{1999AJ....117..621H} was not satisfied. 

Each system was simulated for 10 Gyr, or until (1) a dynamical instability occurred according to the criterion of \citet{1999AJ....117..621H}, or (2) the planet collided with, or was tidally disrupted by the primary star (assuming a tidal disruption radius $r_\mathrm{t} = \eta R_\mathrm{p} \, [M_\star/m_\mathrm{p}]^{1/3}$ with $\eta=2.7$ \citealt{2011ApJ...732...74G}). According to \textsc{SeBa}, the fraction of time of 10 Gyr spent during the various evolutionary stages assuming $M_\star = 1.2 \, M_\odot$ ($M_\star = 6.0 \, M_\odot$) is $\approx 0.56$ ($\approx 0.007$) for the MS, $\approx 0.09$  ($\approx 0.008$) for the giant phases (including core helium burning, i.e. from RGB up to and including AGB), and $\approx 0.35$ ($\approx 0.985$) for the WD phase.

\begin{figure}
\center
\includegraphics[scale = 0.45, trim = 5mm 10mm 0mm 10mm]{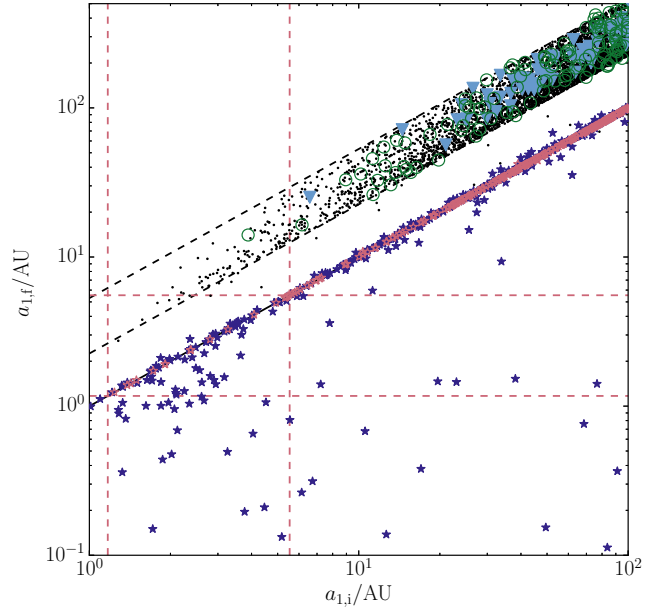}
\caption{\small Initial versus final $a_1$, showing 5 \% of all simulated systems. Refer to \S\,\ref{sect:results:overview} for the meaning of the symbols. Red dashed lines: the maximum radii of the primary star for the lowest and highest masses considered (1.2 and 6 $M_\odot$). Black dashed lines: adiabatic mass loss lines for the mass boundaries. }
\label{fig:initial_final_smas_MC03_lm}
\end{figure}

\section{Results}
\label{sect:results}
\subsection{Overview}
\label{sect:results:overview}

In \F\,\ref{fig:initial_final_smas_MC03_lm}, we show initial versus final $a_1$. The various outcomes are distinguished with symbols and colours, as described below. 

\begin{figure}
\center
\includegraphics[scale = 0.45, trim = 0mm 5mm 0mm 10mm]{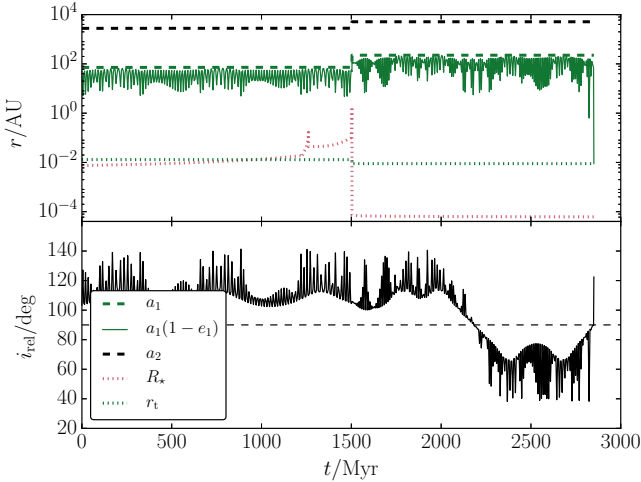}
\caption{\small Example evolution in which the planet is tidally disrupted by the star after the latter has evolved to a WD. Top panel: various distances of interest: the planet's semimajor axis $a_1$ (dashed green line) and periapse distance $a_1(1-e_1)$ (solid green line), the binary orbit semimajor axis $a_2$ (black dashed line), the primary stellar radius $R_\star$ (red dotted line) and the planetary tidal disruption radius $r_\mathrm{t}$ (green dotted line). Bottom panel: the inclination between the planetary and binary companion orbits. The dashed line shows $90^\circ$. The primary star RGB and AGB phases occur near $\approx 1250 \, \mathrm{Myr}$ and $\approx 1500 \, \mathrm{Myr}$, respectively. During the pre-WD phase, the periapse distance $a_1(1-e_1)$ oscillates due to LK cycles, but does not become small enough for strong tidal dissipation, tidal disruption or collision with the primary star. After the AGB phase, the LK eccentricity oscillations increase in amplitude due to the decrease in $a_2/a_1$, with a similar minimum $a_1(1-e_1)$ whereas $a_1$ has increased due to mass loss. At $\approx 2800 \, \mathrm{Myr}$, a flip occurs in the orbital orientation from prograde ($<90^\circ$) to retrograde ($>90^\circ$), which is associated with a very high eccentricity and $a_1(1-e_1) \approx 10^{-2} \, \mathrm{AU}$, triggering the tidal disruption of the planet. }
\label{fig:example_MC03_ex}
\end{figure}

\begin{enumerate}[label=(\roman*)]
\item Black dots in \F\,\ref{fig:initial_final_smas_MC03_lm} -- stable planets in expanded orbits, on lines associated with adiabatic mass loss, $a_{1,\mathrm{f}} = a_\mathrm{1,i} \, (M_{\star,\mathrm{MS}}/M_{\star,\mathrm{WD}})$. Given the range of $M_\star$, this results in a band of systems bounded by the two black dashed adiabatic mass loss lines. 
\item Dark blue filled stars -- pre-WD collisions, on or below $a_{1,\mathrm{f}} = a_\mathrm{1,i}$. After the main-sequence (MS) phase, tidal dissipation becomes more efficient. Possibly coupled with LK cycles, this leads to planetary engulfment.  
\item Light red open stars -- pre-WD tidal disruptions, on $a_{1,\mathrm{f}} = a_\mathrm{1,i}$. The inner orbit eccentricity is excited by LK cycles during the MS. This leads to tidal disruption in a highly eccentric orbit because tidal friction in the radiative envelope is very weak. 
\item Green open circles -- post-WD tidal disruptions, within the same band as (i). After the AGB mass loss phase, the decreased semimajor axis ratio $a_2/a_1$ gives rise to extremely high eccentricities and tidal disruption. An example is given in \F\,\ref{fig:example_MC03_ex}.
\item Blue filled triangles -- dynamically unstable systems (according to the criterion of \citealt{1999AJ....117..621H}), triggered by AGB mass loss. 
\end{enumerate}

\begin{figure}
\center
\includegraphics[scale = 0.43, trim = 0mm 5mm 0mm 10mm]{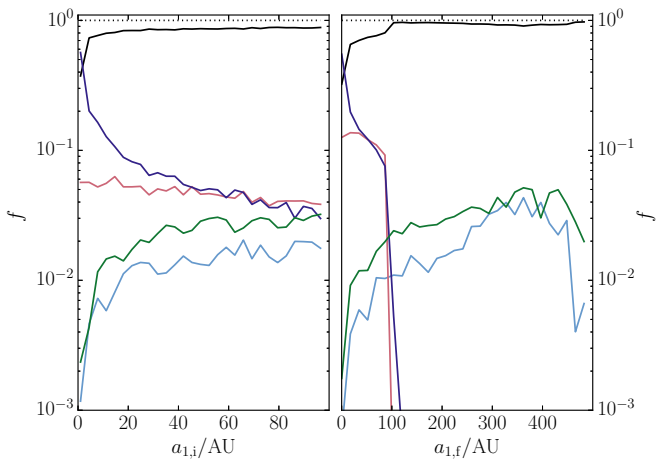}
\caption{\small The fractions of systems corresponding to the outcomes described in \S\,\ref{sect:results:overview} as a function of $a_{1,\mathrm{i}}$ (left panel) and $a_{1,\mathrm{f}}$ (right panel).}
\label{fig:fractions_MC03_lm}
\end{figure}

In \F\,\ref{fig:fractions_MC03_lm}, the fractions of systems corresponding to the outcomes are shown as a function of $a_{1,\mathrm{i}}$ (left panel) and $a_{1,\mathrm{f}}$ (right panel). The fractions for $a_{1,\mathrm{f}}>100 \, \mathrm{AU}$ are incomplete for outcomes (ii) and (iii).

For small $a_{1,\mathrm{i}}$, the fraction of systems with planets being engulfed during the pre-WD phase is unity, and decreases as $a_{1,\mathrm{i}}$ increases. There is a minimum $a_{1,\mathrm{i}}$ for which planets can be tidally disrupted after WD formation, or for which a dynamical instability occurs. From \F\,\ref{fig:fractions_MC03_lm}, this minimum is $a_{1,\mathrm{i}} \gtrsim 5 \, \mathrm{AU}$ (or $a_{1,\mathrm{f}} \gtrsim 10 \, \mathrm{AU}$). Beyond the minimum value, the fraction of post-WD tidally disrupted planets (dynamically unstable systems) is approximately constant at $\sim 0.03$ ($\sim 0.01$).

\begin{figure}
\center
\includegraphics[scale = 0.45, trim = 5mm 5mm 0mm 10mm]{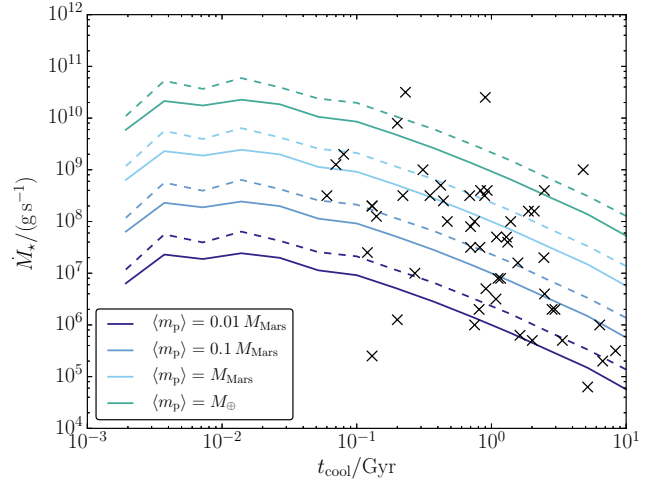}
\caption{\small Simulated WD accretion rates as a function of cooling time (solid lines; dashed lines indicate the standard deviation) assuming various mean planetary masses (indicated in the legend). Black crosses: observational data from \citet{2009ApJ...694..805F}. }
\label{fig:accretion_rates_MC03_lm}
\end{figure}

\subsection{WD pollution -- comparisons to observations}
\label{sect:results:pol}
Outcome (iv) is expected to result in WD pollution. In \F\,\ref{fig:accretion_rates_MC03_lm}, we show WD accretion rates as a function of cooling time from the simulations (solid and dashed lines), and observations (crosses, from \citealt{2009ApJ...694..805F}). Simulated accretion rates were computed from post-WD tidal disruption events assuming that $(1/2)\, m_\mathrm{p}$ is eventually accreted onto the WD \citep{1988Natur.331..687H}. Disruption rates were found to be independent of planetary mass. Using this result, we assumed a range of mean planetary masses $\langle m_\mathrm{p}\rangle$ in \F\,\ref{fig:accretion_rates_MC03_lm}.

Both simulated and observed accretion rates tend to decrease with cooling time. The bulk of the observations can be explained with $\langle m_\mathrm{p}\rangle$ ranging between $\sim 0.01$ and $1\,M_\mathrm{Mars}$. Nongravitational effects may, however, be important for masses $\lesssim 0.3 \, M_\mathrm{Mars}$.

\begin{figure}
\center
\includegraphics[scale = 0.45, trim = 5mm 5mm 0mm 10mm]{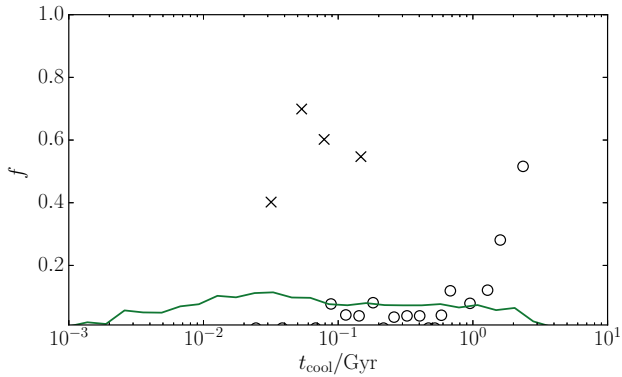}
\caption{\small Solid line: the fraction of polluted WDs as a function of cooling time. Black circles and crosses: observed pollution fractions from \citet{2006A&A...453.1051K} and \citet{2014A&A...566A..34K}, respectively. }
\label{fig:pollution_fractions_MC03_lm}
\end{figure}

In \F\,\ref{fig:pollution_fractions_MC03_lm}, we show the fractions of polluted WDs as a function of cooling time (assuming a binary fraction of 0.5), and including observations from \citet{2014A&A...566A..34K}. For cooling times between $\sim 0.1$ and 1 Gyr, the fractions from the simulations, $\sim 0.05$, are consistent with the observed fractions. The simulations are unable to produce fractions as high as $\sim 0.7$ for cooling times of $\sim 0.05\,\mathrm{Gyr}$, or $\sim 0.5$ for cooling times of $\sim 2 \, \mathrm{Gyr}$.

\section{Discussion}
\label{sect:discussion}
\subsection{Approximations in the dynamics}
\label{sect:summary_discussion:appr_dyn}
In our simulations, the dynamics were modelled using the computationally advantageous secular approach. However, in the `semisecular' regime of $3 \lesssim a_2(1-e_2)/a_1\lesssim 10$ \citep{2012ApJ...757...27A,2014ApJ...781...45A}, in which the system is still dynamically stable, the approximations made in the secular method break down. In our simulations, $\approx 0.5$ of the the tidally disrupted systems have $a_2(1-e_2) > 10$ (at the moment of disruption). For the group in the semisecular regime, we expect that the true eccentricity excitation (i.e. as computed with direct $N$-body integrations) is at least as effective compared to the secular method, if not higher (see e.g. Fig. 5 of \citealt{2014ApJ...781...45A}). Therefore, we do not expect that this strongly affects our conclusions regarding WD pollution. Regarding uncertainties associated with the finite order of the expansion in the secular method, we also carried out the population synthesis up and including third-order terms (by default, terms up to and including fifth order were included), and found no statistically distinguishable results. 

If $a_2(1-e_2)/a_1$ is even smaller, then a short-term dynamical instability can occur. In our simulations, these conditions for dynamical instability are invariably triggered at WD formation (zero cooling ages), and the fraction of systems is lower compared to the `dynamically stable' tidal disruption systems by a factor of a few (cf. \F\,\ref{fig:fractions_MC03_lm}). Such dynamical instabilities can lead to collisions, but also to ejections, most likely of the planet. In the simulations of \citet{2012ApJ...760...99P}, roughly equal-mass stars were considered, and $\approx 0.01$ of the cases led to collisions of objects. Therefore, we do not expect a large contribution to WD pollution from tidal disruptions following a dynamical instability at WD formation.

\section{Conclusions}
\label{sect:conclusions}
We considered a scenario for WD pollution by planets triggered by LK oscillations induced by a binary companion. Our computed accretion rates are consistent with observations for planetary masses between $\sim 0.01$ and $1\,M_\mathrm{Mars}$. The fraction of polluted WDs is consistent with observations of WDs with intermediate cooling times ($0.1 \, \mathrm{Gyr} \lesssim t_\mathrm{cool} \lesssim 1\,\mathrm{Gyr}$). For short and long cooling times, our scenario cannot explain the high observed pollution fractions of up to 70 per cent. Our scenario may also apply to planetesimals, but further work is needed to incorporate nongravitational effects.

\section*{Acknowledgements}
We thank the anonymous referee for useful comments. This work was supported by the Netherlands Research Council NWO (grants \#639.073.803 [VICI],  \#614.061.608 [AMUSE] and \#612.071.305 [LGM]) and the Netherlands Research School for Astronomy (NOVA). 

\bibliographystyle{mnras}
\bibliography{literature}

\label{lastpage}
\end{document}